\documentclass[referee]{raa}

\usepackage{graphicx,times}
\usepackage{natbib}
\usepackage{amssymb,amsmath}
\bibpunct{(}{)}{;}{a}{}{,}

\usepackage[pagebackref=true]{hyperref}

\usepackage[dvipsnames]{xcolor}

\begin{document}

\title{Eleven Local Volume dwarf galaxies in the FASHI survey}
%\subtitle{Subtitle}

\volnopage{ {\bf 20XX} Vol.\ {\bf X} No. {\bf XX}, 000--000}
\setcounter{page}{1}

\author{
Aleksandra Nazarova\inst{1}
\and Dmitry Makarov\inst{1}
\and Igor Karachentsev\inst{1}
\and Chuan-Peng Zhang\inst{2,3}
\and Ming~Zhu\inst{2,3}
}

\institute{
Special Astrophysical Observatory, Russian Academy of Sciences, Nizhnii Arkhyz, 369167 Russia
\and
National Astronomical Observatories, Chinese Academy of Sciences, Beijing, 1000101, PR China
\and
Guizhou Radio Astronomical Observatory, Guizhou University, Guiyang 550000, PR China\\
}

%\date{Received XXX, 2025}

\abstract{
We determined HI parameters for eleven nearby late-type dwarf galaxies using FASHI data cubes, despite the fact that the first version of the FASHI catalog does not list any radio sources that could correspond to these galaxies.
Four of them are probable peripheral satellites of the bright spiral galaxies: NGC\,3556, NGC\,4258, NGC\,4274, NGC\,4490, while others are isolated objects. 
The considered sample has the following median parameters: 
a heliocentric velocity of $V_\mathrm{h} = 542$~km\,s$^{-1}$, 
an HI-line width of $W_\mathrm{50} = 28$~km\,s$^{-1}$, 
an hydrogen mass of $\log(M_\mathrm{HI}/M_\odot)= 6.83$, 
a stellar mass of $\log(M_{*}/M_\odot) = 7.19$, 
and a specific star formation rate of $\mathrm{sSFR} = -10.17$~yr$^{-1}$. 
}

\authorrunning{Nazarova et al.}
\titlerunning{Eleven Local Volume dwarf galaxies in the FASHI survey}

\maketitle

\section{Introduction}

The discrepancies with the predictions of the standard cosmological $\Lambda$CDM modeling are most evident on small scales, when the theory is compared with observations of the most numerous class of galaxies -- dwarf galaxies. A Local Volume (LV) of the Universe with a radius of 12~Mpc around the Milky Way provides a unique opportunity to study such low-mass objects, which are usually inaccessible to observations at larger distances. Most sky surveys are limited by an apparent magnitude or flux of objects. While they are effective at detecting giant galaxies, they are substantially less complete in tracing the dwarf galaxies. Being limited by distance, LV avoids bias toward high surface brightness galaxies and provides a more homogeneous representation of different galaxy populations. 

Currently, the Local Volume (LV) galaxy database, available at http://www.sao.ru/lv/lvgdb, contains more than 1500 galaxies. 
About 85\% of them are dwarf galaxies with stellar masses smaller than $10^9$~M$_\odot$~\citep{2019AstBu..74..111K}.
A systematic search for new dwarf galaxies is required to improve the completeness of the data in the low-mass end.
In recent years, thanks to the deep images provided by the Dark Energy Spectroscopic Instrument (DESI) Legacy Imaging Surveys~\citep{2019AJ....157..168D}, more than a hundred candidates for the LV dwarf galaxies have been discovered~\citep{2022AstBu..77..372K, 2023A&A...678A..16K, 2023Ap.....66..441K, 2025PASA...42...e026}.

Large-scale extragalactic HI surveys such as the HI Parkes All Sky Survey~\citep{1997ApJ...490..173Z}, the Arecibo Legacy Fast ALFA~\citep{2005AJ....130.2598G}, FAST All Sky HI~\citep[FASHI]{2024SCPMA..6719511Z} are essential for contribution to the population of dwarf galaxies in the Local Universe. 
They provide measurements that determine the shape of the HI mass function~\citep{2010ApJ...723.1359M, 10.1093/mnras/sty521, 2025A&A...695A.241M}, and its variation with the environment~\citep{10.1093/mnras/stw263, 10.1093/mnras/sty521}. 
These surveys have the potential to discover objects with unusual properties, such as nearby faint dwarf Leo\,P~\citep{2013AJ....146...15G}, 
gas-rich halo around ultra-faint dwarf KK\,153~\citep{2025ApJ...982L..36X}
almost dark galaxies~\citep{2015ApJ...801...96J, 2023ApJ...944L..40X}, 
high-velocity cloud as a potential dark galaxy~\citep{2025SciA...11S4057L},
and, in contrast, dwarf galaxies with a dark matter deficit~\citep{2020NatAs...4..246G}. 
The study of HI content in dwarf galaxies, particularly at the faint, low-mass end of the luminosity function, plays a crucial role in understanding galaxy formation and evolution.
It provides fundamental insights into the kinematics of HI disks in low-mass systems~\citep{2015AJ....149..180O, 2016AJ....152..157L},
the structural parameters of the gas distribution~\citep{2008MNRAS.386.1667B, 2023PASA...40...32R},
the relative contributions of baryions (gas and stars)~\citep{2006ApJ...653..240G, 2024ARA&A..62..113H}
and connection with dark matter~\citep{2017MNRAS.467.2019R, 2025A&A...699A.311M},
and the physical processes governing gas accretion and loss~\citep{2012ARA&A..50..491P},
emphasizing the important and universal role of halo spin in galaxy evolution and formation~\citep{2025RAA....25a1001R}.
Furthermore, it is essential for examining the baryonic Tully–Fisher relation~\citep{2000ApJ...533L..99M, 2016ApJ...816L..14L} 
and extend it to very low-mass galaxies~\citep{2023ApJ...947L...9H, 2024A&A...689L...3L, 2025MNRAS.538..775H}.
Moreover, examining how the mass function and gas content of dwarf galaxies vary with environment and large-scale structure is essential for linking small-scale processes to the cosmological framework~\citep{2009A&A...498..407G, 2022MNRAS.510.1716R, 2024MNRAS.528.4139H}.
Also, modern observational capabilities allows researchers to detect gas not only by HI emission, but also by HI absorption~\citep{2025ApJS..276....6Z}, 
as well as to track the evolution of the dwarf galaxy properties with redshift~\citep{2022ApJ...940L..10B}.

The identifications of radio sources from blind HI surveys with optical counterparts~\citep{2008AstL...34..832K}, accompanied by distance estimates, allows us to expand the LV sample by discovering new galaxies and measuring the velocities of already known objects.
In a recent paper, \citet{2024A&A...684L..24K} found 20 new LV galaxies identified as optical counterparts to FASHI sources. 
Unfortunately, LV covers a relatively small region of space and the sample properties are subject to cosmic variations.
Increasing the depth of the distance-limited sample is very important for creating a representative sample of galaxies. 
By extending the search for nearby galaxies using the FASHI survey to a distance of about 15~Mpc, we have discovered about 60 new dwarf galaxies (Nazarova et al. 2025, paper in preparation) in addition to those discovered by \citet{2024A&A...684L..24K}.

Typically, dwarf galaxies have weak HI signals with narrow HI widths, requiring higher sensitivity with smaller velocity bins. The FASHI survey is well suited for this purpose. It has a higher spectral and spatial resolution and covers a larger sky area than the ALFALFA survey, allowing it to sample a cosmologically more fair volume. 
According to~\citet{2024SCPMA..6719511Z}, the FAST All Sky HI Survey (FASHI) will cover the entire sky observable by the Five-hundred-meter Aperture Spherical radio Telescope (FAST), spanning approximately 22000 square degrees of declination between $-14^\circ$ and $+66^\circ$, and in the frequency range of 1050--1450~MHz. It achieves a median detection sensitivity of $\sim$0.76~mJy\,beam$^{-1}$, a spectral line velocity resolution of $\sim$6.4~km\,s$^{-1}$ at a frequency of $\sim$1.4~GHz and a spatial resolution of $2.9^\prime$ at 1420~MHz. 
The first release of the FAST survey, with data collected between August 2020 and June 2023, covers more than 7600 square degrees and contains 41741 extragalactic HI sources in the heliocentric redshift range $200<cz_\odot<26323$~km\,s$^{-1}$. When completed, FASHI is expected to be the largest extragalactic HI catalog.

In this study, we evaluated the potential of the blind FASHI survey~\citep{2024SCPMA..6719511Z} to detect HI emission from faint dwarf galaxies.
Our approach involved a detailed examination of regions around selected nearby irregular galaxies where the presence of gas is expected but has not been previously detected. 
If effective, this method could be applied to a broad population of dwarf galaxies, significantly extending the HI mass and line-width functions into the low-mass tail. 

\section{Sample selection}

\begin{figure*}
\centering
\includegraphics[width=\textwidth]{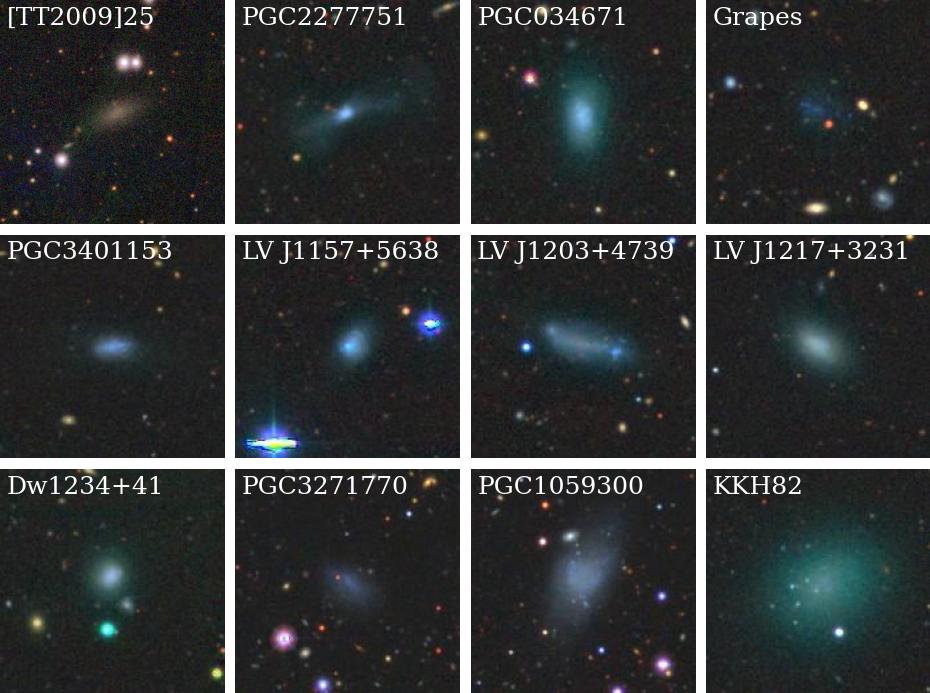}
\caption{
Optical images of the 12 nearby galaxies from the DESI Legacy Imaging Surveys (\textrm{[TT2009]\,25} from Pan-STARRS1). 
Each image covers $2\arcmin\times2\arcmin$; North is up and East is to the left.
}
\label{fig:mosaic}
\end{figure*}

During cross-matching LV galaxies falling within the FASHI survey footprint with the list of sources from the FASHI catalog~\citep{2024SCPMA..6719511Z}, we found that some dwarf irregular galaxies have no radio counterparts. 
This result was not unexpected, given that the FASHI catalog only includes sources of relatively high confidence, excluding those located near the edges of the data cubes or those with a low signal-to-noise ratio. %($< 5.0$). 
The first FASHI catalog also shows uneven detection sensitivity in different regions. 
This may have led to a lack of faint radio sources in regions with the lowest sensitivity.
In addition, automatic object extraction may overlook faint objects in complicated fields.

To assess the feasibility of detecting signals and measuring the basic HI properties of faint galaxies using a manual rather than automated approach, we selected 11 late-type dwarf galaxies located within the FASHI survey area for which no corresponding radio sources are present in the FASHI catalog.
To simplify the task, we selected galaxies with known optical radial velocities that lie within the FASHI velocity range and are sufficiently isolated to ensure that their radio signals would not overlap with those of nearby galaxies.
The galaxies in our sample have not previously been observed in HI. 
Nevertheless, their blue stellar populations indicate recent star formation, and therefore a substantial amount of neutral hydrogen is expected. 
In other words, based on these criteria, we selected the 11 most promising dwarf galaxies for radio detection.
Images of the selected galaxies from the DESI Legacy Imaging Surveys are presented in Fig.~\ref{fig:mosaic}, with the addition of a 12th galaxy, PGC~3271770 (see discussion below). As can be seen in Fig.~\ref{fig:mosaic}, the optical diameters of the galaxies are smaller than the beam size of $\mathrm{FWHM} = 2.9^\prime$ and do not exceed $1^\prime$.
We performed aperture photometry in the $g$ and $r$-bands for all of these galaxies, constructed growth curves, and determined the asymptotic total magnitudes using publicly available data from the DESI Legacy Surveys.

\section{Results}

We extracted 3D HI data cubes centered on the optical coordinates of these galaxies with an average angular size approximately $66^\prime \times 50^\prime$.
Each cube covers a velocity range of 150.0--2000.0~km\,s$^{-1}$ taken in the radio convention.
To convert velocities to the optical convention $cz$, we used the equation 
\begin{equation}
cz = \frac{v_\mathrm{rad}}{1 - v_\mathrm{rad}/c},
\end{equation}
where $v_\mathrm{rad}$ is the velocity in the radio convention and $c$ is the speed of light. 
In our case, the typical correction was about 0.17\%.

An initial inspection of the HI maps revealed a noticeable signal in roughly half of our targets. 
Next, we carried out a detailed analysis of the spectra.
The known positions and optical velocities of the galaxies were crucial for identifying the HI emission. 
Careful selection of the spatial and spectral regions around the line within the data cubes subsequently allowed us to extract HI spectra for all galaxies, with the only exception of [TT2009]\,25, where no emission was detected.
In addition, at the periphery of the field centered on PGC\,1059300, we detected a reliable HI signal associated with a more distant and fainter dwarf galaxy, PGC\,3271770.

For each detected signal, a region containing the entire flux from the galaxy was extracted. Such a subcube, with a typical size of $10^\prime \times 6^\prime$, was then integrated into a 1D spectrum. The HI profiles and 2D~0-moment~maps derived from the FASHI data cubes, along with the corresponding optical images from the DESI Legacy Imaging Surveys, are presented in Appendix~\ref{sec:HIcubes}.
The galaxies in our sample are dwarf systems, which typically exhibit HI line widths of approximately 30~km\,s$^{-1}$ and can be well represented by a single Gaussian profile. 
By fitting the HI line with a Gaussian, we derived the redshift and obtained an estimate of $W_{50}$, defined at the 50\% level of the line peak. 
The total flux was measured by summing all velocity channels containing signal within the appropriate velocity range.

The parameters of the studied galaxies are summarized in Table~\ref{tab:HIparameters} containing: 
(1) galaxy name as given in the Updated Nearby Galaxy Catalog~\citep[UNGC,][]{2013AJ....145..101K};
(2) optical equatorial coordinates;
(3) optical heliocentric velocity (in km\,s$^{-1}$);
(4--6) HI parameters derived from the FASHI cubes: 
integrated HI flux (in Jy\,km\,s$^{-1}$), 
velocity in the optical convention (in km\,s$^{-1}$), 
and HI line width at half maximum intensity (in km\,s$^{-1}$);
(7--8) apparent total $g$ and $r$ magnitudes measured by us using images from the DESI Legacy Imaging Surveys\footnote{\url{https://www.legacysurvey.org/}}~\citep{2019AJ....157..168D}; 
(9) apparent total $B$-band magnitude, estimated from the $g$ magnitude and $(g - r)$ color using the transformation equations from The Dark Energy Survey~\citep{2021ApJS..255...20A}; 
(10) $m_{21}$ magnitude, calculated as $m_{21}= 17.4 - 2.5 \log(S_{\rm HI})$ according to the formula adopted in the Third Reference Catalogue of Bright Galaxies~\citep[RC3,][]{1991rc3..book.....D} and in the HyperLEDA database\footnote{\url{http://leda.univ-lyon1.fr/}}~\citep{2014A&A...570A..13M}.

\setlength{\tabcolsep}{1.8pt}
\begin{table*}
\centering
\caption{HI parameters derived from the FASHI cubes for the galaxies in our sample.} 
\label{tab:HIparameters}
\begin{tabular}{l c l c r@{ }l l c c c c} 
\hline\hline
 &  
 & 
 & 
\multicolumn{4}{c}{FASHI HI data} &
 &
 &
 &
\\

\cline{4-7}

Name &  
RA(2000.0)Dec  & 
\multicolumn{1}{c}{$V_\mathrm{h}^{(a)}$} & 
\multicolumn{1}{c}{$S_\mathrm{HI}$} &
\multicolumn{2}{c}{$cz_\odot$} &
\multicolumn{1}{c}{$W_\mathrm{50}$} &
$g$ &
$r$ &
$B$ &
$m_{21}$ \\

 &
 &
\multicolumn{1}{c}{\footnotesize{km\,s$^{-1}$}} &
\multicolumn{1}{c}{\footnotesize{Jy\,km\,s$^{-1}$}} &
\multicolumn{2}{c}{\footnotesize{km\,s$^{-1}$}} &
\multicolumn{1}{c}{\footnotesize{km\,s$^{-1}$}} &
\footnotesize{mag} &
\footnotesize{mag} &
\footnotesize{mag} &
\footnotesize{mag} \\

\hline

% Name                            RA DEC               Vh            S_HI                 cz                  W50          g       r       B      m21
\lbrack TT2009\rbrack\,25   & 022112.4+422151   & $692\pm58$  &                 &          &            &               &       &       & 17.90\rlap{$^{(a)}$} &       \\
PGC2277751                  & 103512.1+461412   & $504\pm23$  & $0.15\pm0.07$   & $542.2$  & $\pm$ 6.0  & $30.3\pm14.1$ & 16.75 & 16.33 & 17.15 & 19.46 \\
PGC034671                   & 111948.6+554322   & $605\pm7$   & $0.58\pm0.03$   & $609.6$  & $\pm$ 0.7  & $31.9\pm1.6$  & 16.01 & 15.52 & 16.45 & 17.99 \\
Grapes                      & 115205.6+544732   & $321\pm210$ & $0.77\pm0.01$   & $223.2$  & $\pm$ 0.1  & $20.9\pm0.2$  & 17.95 & 17.72 & 18.25 & 17.68 \\
PGC3401153                  & 115352.4+512938   & $497\pm42$  & $0.34\pm0.01$   & $513.7$  & $\pm$ 0.2  & $25.7\pm0.5$  & 17.17 & 16.80 & 17.54 & 18.57 \\
LV\,J1157+5638              & 115754.2+563817   & $416\pm1$   & $0.45\pm0.03$   & $404.2$  & $\pm$ 0.9  & $30.4\pm2.0$  & 16.80 & 16.53 & 17.12 & 18.27 \\
LV\,J1203+4739              & 120300.0+473915   & $492\pm14$  & $1.40\pm0.10$   & $499.3$  & $\pm$ 1.8  & $58.3\pm4.2$  & 16.37 & 15.99 & 16.75 & 17.03 \\
LV\,J1217+3231              & 121732.0+323157   & $447\pm10$  & $0.18\pm0.02$   & $445.2$  & $\pm$ 1.6  & $27.8\pm3.7$  & 16.43 & 15.92 & 16.85 & 19.26 \\
Dw1234+41                   & 123438.2+411634   & $614\pm6$   & $0.133\pm0.002$ & $610.1$  & $\pm$ 0.6  & $24.3\pm0.3$  & 16.79 & 16.21 & 17.28 & 19.62 \\
PGC1059300                  & 130000.3$-$042138 & $540\pm89$  & $2.42\pm0.29$   & $796.5$  & $\pm$ 3.5  & $69.4\pm8.2$  & 16.02 & 15.65 & 16.36 & 16.44 \\
KKH82                       & 131258.7+414712   & $529\pm40$  & $0.190\pm0.004$ & $547.8$  & $\pm$ 0.2  & $17.2\pm0.4$  & 15.31 & 14.69 & 15.82 & 19.20 \\
\hline
PGC3271770$^{(b)}$          & 125959.0$-$042457 &             & $0.51\pm0.05$   & $1447.5$ & $\pm$ 1.1  & $26.9\pm2.5$  & 17.80 & 17.54 & 18.08 & 18.13 \\
\hline\hline
\end{tabular}
\tablecomments{0.86\textwidth}{$^{(a)}$Taken from LV database~\citep{2012AstBu..67..115K}. $^{(b)}$From PGC1059300's field.}
\end{table*}

%In the following, we provide brief comments on several objects.

\begin{description}

\item[\textit{[TT2009]\,25}:]
This transition-type dwarf, located at a distance of $10.3^{+1.2}_{-1.7}$~Mpc~\citep{2019A&A...629L...2M}, is a satellite of NGC\,891.
Its FUV emission indicates recent star formation, but the absence of H$\alpha$~\citep{2015Ap.....58..453K} suggests that star formation ended more than 10~Myr ago.
We did not detect an HI signal in the FASHI data.
This is consistent with the idea that the system depleted its hydrogen reservoir during the most recent episode of star formation.

\item[\textit{PGC\,2277751}:] 
This isolated dwarf galaxy, located at a distance of $D = 11.48$~Mpc~\citep{2022AstBu..77..388T}, exhibits clear signs of a recent merger, and its weak HI signal hints at deviations from a Gaussian profile.
We also observed this galaxy with the 100-meter Robert C.~Byrd Green Bank Telescope\footnote{The National Radio Astronomy Observatory and Green Bank Observatory are facilities of the U.S. National Science Foundation operated under cooperative agreement by Associated Universities, Inc.} (GBT) under project GBT/24B--157 (P.I.\ Nazarova), and estimated the following parameters:
$S_\mathrm{HI} = 0.26\pm0.03$~Jy\,km\,s$^{-1}$, $V_\mathrm{h} = 544.0\pm1.9$~km\,s$^{-1}$, $W_\mathrm{50} = 43.9\pm4.5$~km\,s$^{-1}$, which are in good agreement with the values obtained in the present study.

\item[\textit{PGC\,034671}] is a satellite of NGC\,3556.

\item[\textit{Grapes}] is an isolated dwarf irregular galaxy with a highly clumpy structure and a very blue color, $(g-r)_0=0.22$.
Its Tip of the Red Giant Branch (TRGB) distance is $D = 5.96$~Mpc~\citep{2021AJ....162...80A}.
It has the highest hydrogen-to-stellar mass ratio of $M_\mathrm{HI}/M_\star = 5.13$ in our sample.

% \item[PGC\,3401153:] 

\item[\textit{LV\,J1157+5638}:] 
In this isolated dwarf galaxy, located at a distance of $D = 8.75^{+0.67}_{-0.62}$~Mpc~\citep{2021AJ....162...80A}, undergoes a modest star formation episode.
About $1.5^\prime$ to the south, there is an extremely low-luminosity galaxy, LV\,J1157+56\,sat ($g = 20.64$, $r = 20.59$, $z = 20.52$) at J115753.06+563649.3, which may be a dwarf satellite of LV\,J1157+5638.

\item[\textit{LV\,J1203+4739}] is a satellite of NGC\,4258. 
The HI profile shows a slight but noticeable deviation from a Gaussian shape.

\item[\textit{LV\,J1217+3231}] belongs to a group of the E-type galaxy NGC\,4274.
Using the Hubble Space Telescope, \citet{2018ApJ...858...62K} were only able to put a lower limit on its distance of $D>13$~Mpc.
This group is part of the Coma\,I cloud, which exhibits enormous negative peculiar velocities of the order of $-1000$~km\,s$^{-1}$~\citep{2011ApJ...743..123K}.
Therefore, it is not surprising that the Numerical Action Method~\citep[NAM,][]{2017ApJ...850..207S,2020AJ....159...67K} gives an implausibly short distance of 5.30~Mpc.

\item[\textit{Dw\,1234+41}:]
This blue compact dwarf is a satellite of the interacting galaxy pair NGC\,4490/85 ($\approx50^\prime$ to NW).
The dwarf itself has a probable dwarf satellite ($g=19.09$, $r=18.44$, $z=18.23$) that locates $18^{\prime\prime}$ to the SW, at J123437.5+411621.
We also observed Dw\,1234+41 with the GBT under project GBT/23B--032 (P.I.\ Cannon) and estimated the following parameters: $S_\mathrm{HI} = 0.19\pm0.03$ Jy\,km\,s$^{-1}$, $V_\mathrm{h} = 614.3\pm1.4$ km\,s$^{-1}$, $W_\mathrm{50} = 28.1\pm3.4$ km\,s$^{-1}$. 
The HI line has a rather complex shape, which leads to difference in the width estimates.

\item[\textit{PGC\,1059300}:] The shape of the HI line is noticeably different from the Gaussian profile.
%Reliable signal, spatially close to the PGC3271770 (see below). The shape of the line is noticeably different from the Gaussian.

\item[\textit{PGC\,3271770}:] 
This object was not included in the original target list. 
Its signal was detected during the processing of the PGC\,1059300 field. The HI line is clearly visible and exhibits a Gaussian profile. This galaxy is listed in the FASHI catalog under ID = 20230001646, with measured parameters $cz_\odot = 1448.10 \pm 1.63$ km\,s$^{-1}$ and $W_\mathrm{50} = 29.60 \pm 3.25 $ km\,s$^{-1}$, which are in good agreement with our measurements.

\item[\textit{KKH\,82}:] 
This transitional type dwarf galaxy locates at a periphery of M\,51 group.
Its surface brightness fluctuation (SBF) distance is $D=7.58^{+0.59}_{-0.65}$~Mpc~\citep{2022ApJ...933...47C}.

\end{description}

\section{Concluding remarks}

\begin{figure}
\centering
\includegraphics[width=\linewidth]{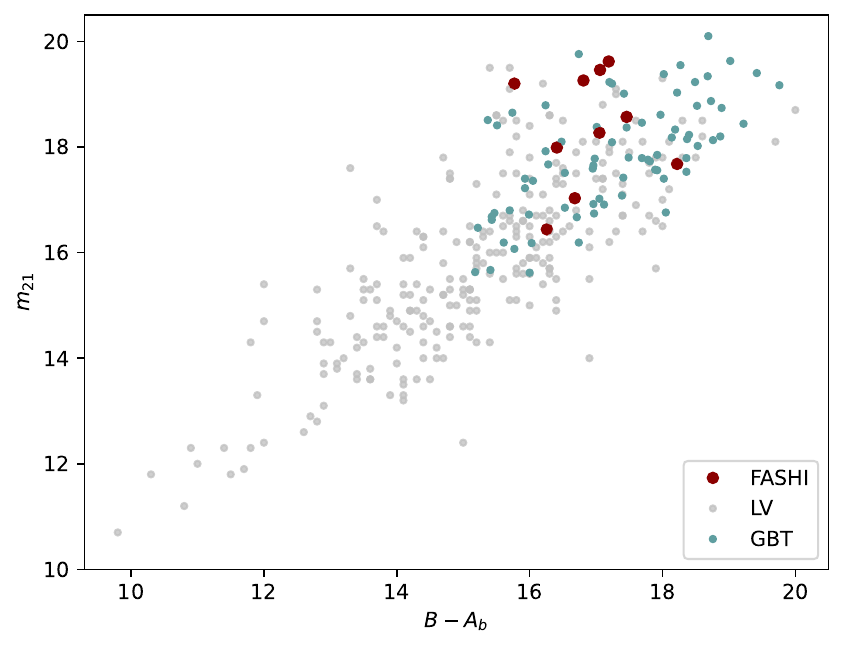}
\caption{
Distribution of 10 studied galaxies (dark red dots) according to $m_{21}$ magnitude and $B$-magnitude corrected for Galactic extinction combined with 77 GBT-detected dwarf galaxies (pale blue dots) and 266 late-type ($T=9$, 10) dwarf galaxies in the Local Volume with TRGB distances (gray dots).
}
\label{fig:B_m21}
\end{figure}

\begin{table}
\centering
\caption{Physical properties of the galaxies} \label{tab:integral_properties}
\begin{tabular}{l r c c c c c } % r r} 
\hline\hline

Name &  
\multicolumn{1}{c}{$D_\mathrm{NAM}$} & 
\multicolumn{1}{c}{$\log M_\mathrm{HI}$} &
\multicolumn{1}{c}{$\log M_\star$} &
\multicolumn{1}{c}{$\log\mathrm{SFR}$} &
\multicolumn{1}{c}{$\log\mathrm{sSFR}$} &
$(g - r)_0$  \\ %&
% $\Theta$\tablefootmark{a} &
% MD \\

 &
\multicolumn{1}{c}{Mpc} &
\multicolumn{1}{c}{$M_\odot$} &
\multicolumn{1}{c}{$M_\odot$} &
\multicolumn{1}{c}{$M_\odot$ yr$^{-1}$} &
\multicolumn{1}{c}{yr$^{-1}$} &
mag \\ % &
% &
% \\

\hline

% Name           D_NAM   M_HI    M_*     SFR       sSFR     g-r         theta     MD
PGC2277751	   & 10.72 & 6.61 & 7.34 & $-2.81$ & $-10.15$ & 0.40 \\ % & $-0.7$ & isolated \\
PGC034671      & 11.79 & 7.28 & 7.80 & $-2.66$ & $-10.46$ & 0.48 \\ % &  1.6   & NGC\,3556  \\
Grapes         &  5.80 & 6.78 & 6.07 & $-4.12$ & $-10.20$ & 0.22 \\ % & $-0.9$ & isolated \\
PGC3401153     &  9.38 & 6.85 & 6.99 & $-2.90$ &  $-9.88$ & 0.35 \\ % & $-0.7$ & isolated \\
LV\,J1157+5638 &  7.84 & 6.81 & 6.85 & $-2.77$ &  $-9.62$ & 0.25 \\ % & $-0.8$ & isolated \\
LV\,J1203+4739 &  8.11 & 7.33 & 7.19 & $-2.77$ &  $-9.96$ & 0.37 \\ % &  0.9   & NGC4258  \\
%LV\,J1217+3231 &  5.30 & 6.07 & 6.97 &         &           & 0.50 \\ % &  0.1   & NGC\,4274  \\
Dw1234+41      &  8.80 & 6.37 & 7.35 & 	       &          & 0.56 \\ % &  0.3   & NGC\,4490  \\
PGC1059300     &  6.72 & 7.41 & 7.16 & $-3.24$ & $-10.41$ & 0.34 \\ % & $-1.2$ & isolated \\
KKH82          &  7.98 & 6.45 & 7.91 & $-3.57$ & $-11.48$ & 0.61 \\ % & $-0.1$ & isolated \\

\hline\hline
\end{tabular}
% \tablecomments{0.86\textwidth}{
% Columns $M_\mathrm{HI}$, $M_\star$, $\mathrm{SFR}$, $\mathrm{sSFR}$ give base-10 logarithmic values: $\log M_\mathrm{HI}$, $\log M_\star$, $\log \mathrm{SFR}$, and $\log \mathrm{sSFR}$.
% }
\end{table}

We used FASHI survey~\citep{2024SCPMA..6719511Z} data to search for the HI signal in 11 LV galaxies with known optical velocities.
By selection criteria, our sample consists of faint dwarfs with $B \gtrsim 16$~mag.
None of them has HI counterparts in the first release of the FASHI catalog. 
Despite the fact that the galaxies were not detected automatically, we tried to extract the signal manually using the corresponding FASHI survey data cubes.
In 10 of the 11 cases, we successfully detected the HI signal and were able to reliably measure the velocity, line width, and flux.
In some cases, the detection was possible solely due to the known optical coordinates and velocities of the galaxies.
Additionally, we measured the HI parameters of a galaxy that accidentally get into one of the data cubes under consideration.
For all galaxies studied we performed surface photometry and determined their total $g$ and $r$ magnitudes using images from the DESI Legacy surveys~\citep{2019AJ....157..168D}.

A comparison of our FASHI-based measurements with optical velocities from the LV database~\citep{2012AstBu..67..115K} shows reasonable agreement within the uncertainties
$\langle V^\mathrm{UNGC} - V^\mathrm{FASHI}\rangle = 23\pm28$~km\,s$^{-1}$ with $\sigma = 85$~km\,s$^{-1}$.
The large scatter is due to the low accuracy of the optical measurements.

Figure~\ref{fig:B_m21} presents the relationship between the radio $m_{21}$ and total optical $B$-band magnitudes.
Our sample is shown in comparison with LV late-type dwarf galaxies that have high-precision distance estimates, excluding those located in regions of high extinction ($A_b > 2\fm0$) or affected by Galactic HI contamination.
We also plot 77 dwarf galaxies recently detected as part of the GBT/23B--032 (P.I.~Cannon) and GBT/24B--157 (P.I.~Nazarova) projects on ``Newly Discovered Local Volume Dwarf Galaxies''~\citep{Nazarova_2025}.
The galaxies in our sample follow the general trend, reflecting the relationship between stellar and hydrogen masses in dwarf galaxies.
However, two-thirds of them lie near the HI flux limit, indicating that the FASHI data enable reliable measurement of the HI signal as low as $S_\mathrm{HI} \approx 0.13$~Jy\,km\,s$^{-1}$.

Table~\ref{tab:integral_properties} summarizes the physical characteristics of the galaxies under study, excluding LV\,J1217+3231, for which a reliable distance estimate could not be obtained.
It contains the following global parameters:
(1) galaxy name;
(2) kinematic distance (in Mpc), determined from the measured radial velocities taking into account local cosmic flows using the Numerical Action Method~\citep[NAM,][]{2017ApJ...850..207S} and a distance-velocity calculator~\citep{2020AJ....159...67K};
(3) neutral hydrogen mass (in solar masses), calculated as $M_{\rm HI}/M_{\odot} = 2.36 \times 10^5 D_{\rm Mpc}^2 S_{\rm HI}$~\citep{2005AJ....130.2598G};
(4) stellar mass (in solar masses), estimated as $\log(M_\star/M_{\odot}) = 11.27 + 2 \log(D_{\rm Mpc}) + 1.45(B - V) - 0.4B$, where $B$ and $V$ magnitudes are corrected for the Galactic extinction. 
This expression is based on the relation $\log(M_\star/L_B) = -0.91 + 1.45(B - V)$~\citep{2016AJ....152..177H}, and incorporates the definition $\log(L_B/L_{\odot})= 0.4 (M_{B,\odot} - M_B)$, with $M_{B,\odot} = 5.44$~mag adopted from \citet{2018ApJS..236...47W};
%stellar mass (in solar masses), estimated using the relation $\log(M_*/L_B) = -0.91 + 1.45(B - V)$, as justified by \citet{2016AJ....152..177H}, where $B$ and $V$ magnitudes are corrected for the Galactic extinction. The $B$-band luminosity is $\log(L_B/L_{\odot})= 0.4 (M_{B,\odot} - M_B)$. Adopting $M_{B,\odot} = 5.44$~\citep{2018ApJS..236...47W}, we finally have $\log(M_*/M_{\odot}) = 11.27 + 2 \log(D_{\rm Mpc}) + 1.45(B - V) - 0.4B$;
(5) integral star-formation rate, SFR (in $M_\odot$~yr$^{-1}$), taken from the LV database\footnote{\url{https://www.sao.ru/lv/lvgdb/}}~\citep{2012AstBu..67..115K};
(6) specific star-formation rate, sSFR (in yr$^{-1}$), estimated as $\mathrm{SFR}/M_\star$;
(7) $(g-r)_0$ color, corrected for Galactic extinction.
%(8) tidal index, determined via distance and mass of the nearest significant neighbor;
%(9) name of the "main disturber"(=MD), i.e. the neighboring galaxy, producing the maximal tidal influence on a galaxy.

Our sample is characterized by a rather narrow range of stellar masses. 
With the exception of Grapes ($\log M_\star = 6.07$), the galaxies span the range $\log M_\star = [6.85$--$7.91]$, with a median of 7.19.
The measured HI masses also fall within a relatively narrow range, $\log M_\mathrm{HI} = [6.37$--$7.41]$, with a median of 6.81.
The corresponding median HI-to-stellar mass ratio is $M_\mathrm{HI}/M_\star=0.51$, which is representative of LV dwarf galaxies~\citep{2018MNRAS.479.4136K}.
The median star formation rate is $\log\mathrm{SFR} = -2.85$.
KKH\,82 and Grapes exhibit significantly lower values of $\log\mathrm{SFR} = -3.57$ and $-4.12$, respectively.
Ironically, these galaxies lie at opposite ends of the hydrogen abundance distribution.
While KKH\,82 (transition-type dwarf) has nearly exhausted its hydrogen reserves through star formation ($M_\mathrm{HI}/M_\star = 0.035$), Grapes (dIrr) exhibits a hydrogen-to-stellar mass ratio $M_\mathrm{HI}/M_\star = 5.13$ close to the maximal values.
At the same time, the specific star formation rate of Grapes, $\log\mathrm{sSFR} = -10.20$, does not differ from the sample median of $\log\mathrm{sSFR} = -10.17$, with a standard deviation of 0.20.
KKH\,82, as expected, shows a significantly lower value, $\log\mathrm{sSFR} = -11.48$.
Note that the median $\log\mathrm{sSFR} = -10.17$ is almost identical to the inverse age of the Universe, with $T_0=13.8$~Gyr~\citep{2020A&A...641A...6P} and  $\log(1/T_0) \approx -10.14$, indicating that the dwarf galaxies in our sample are reproducing their stellar mass at the current star formation rate over a Hubble time.

\begin{figure}
\centering
\includegraphics[width=\linewidth]{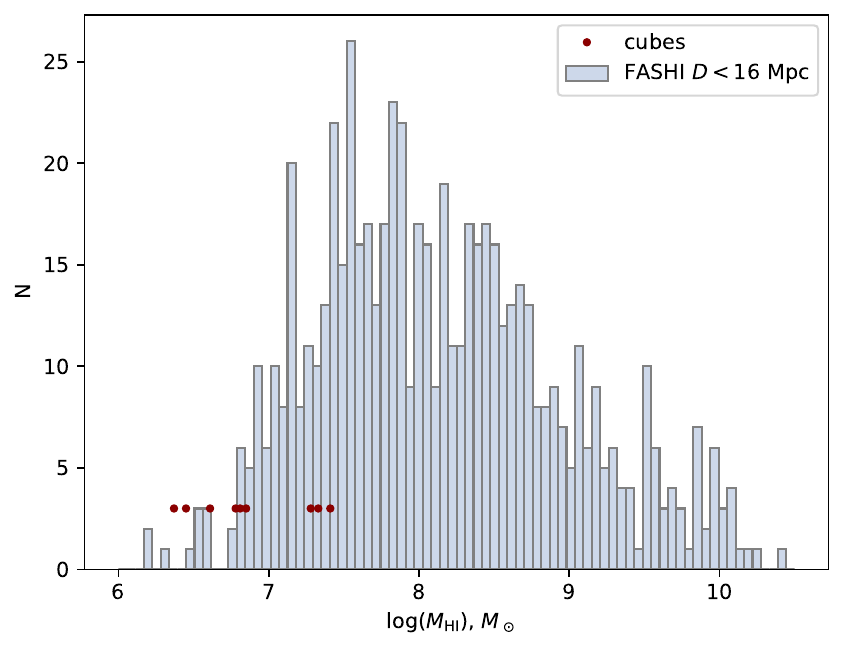}
\caption{
Comparison of the HI masses of galaxies in our sample with the HI mass function of nearby FASHI galaxies with $V_\mathrm{LG}\leq1150$~km\,s$^{-1}$.
}
\label{fig:hist_M_HI}
\end{figure}

Figure~\ref{fig:hist_M_HI} presents the HI mass function of nearby galaxies from the FASHI survey with $V_\mathrm{LG}\leq1150$~km\,s$^{-1}$, in the Local Group centroid reference frame~\citep{1996AJ....111..794K}.
This velocity corresponds to a distance of about 16~Mpc.
The galaxies in our sample are overlaid on the histogram, reflecting that most of them are near the lowest detected masses, where significant incompleteness in the data is observed.
This example demonstrates that searching for a HI-signal in known nearby dwarf galaxies can substantially improve the statistics in the critically important region at the lower end of the HI mass function.

This study demonstrates the potential for detecting HI signals in cases where automatic methods fail.
Extending this research to a larger sample of galaxies is crucial for measuring their redshifts, probing internal kinematics, and estimating gas content.
This is extremely important for building a sample of the LV galaxies limited by distance and ensuring its completeness.
A representative sample of nearby galaxies offers valuable insight into their formation and evolution, spatial distribution, and connection to the large-scale structure of the Universe.
Analyzing the positions and peculiar motions of galaxies in the nearby Universe enables the construction of a detailed map of matter distribution, the tracing of variations in galaxy properties with respect to their location in the cosmic web, and the testing of theoretical models.
Expanding the sample with additional dwarf galaxies will allow for a reliable quantification of the low-mass end of the HI mass function.

\normalem
\begin{acknowledgements}
This work was supported by the Russian Science Foundation grant \textnumero~24--12--00277. %, \url{https://rscf.ru/project/24-12-00277}.
We acknowledge the usage of the HyperLeda database\footnote{\url{http://leda.univ-lyon1.fr}}~\citep{2014A&A...570A..13M}.
The database of galaxies in the Local Volume\footnote{\url{https://www.sao.ru/lv/lvgdb/}}~\citep{2012AstBu..67..115K} was also used in this work.

%We acknowledge the usage of the HyperLeda database\footnote{\url{http://leda.univ-lyon1.fr}}~\citep{2014A&A...570A..13M}.
\end{acknowledgements}

\bibliographystyle{raa}
\bibliography{art}

\begin{appendix}

\onecolumn

\section{HI profiles, HI maps and the corresponding areas on the optical images.}
\label{sec:HIcubes}

\begin{figure*}[h!]
\centering
    \includegraphics[width=\textwidth]{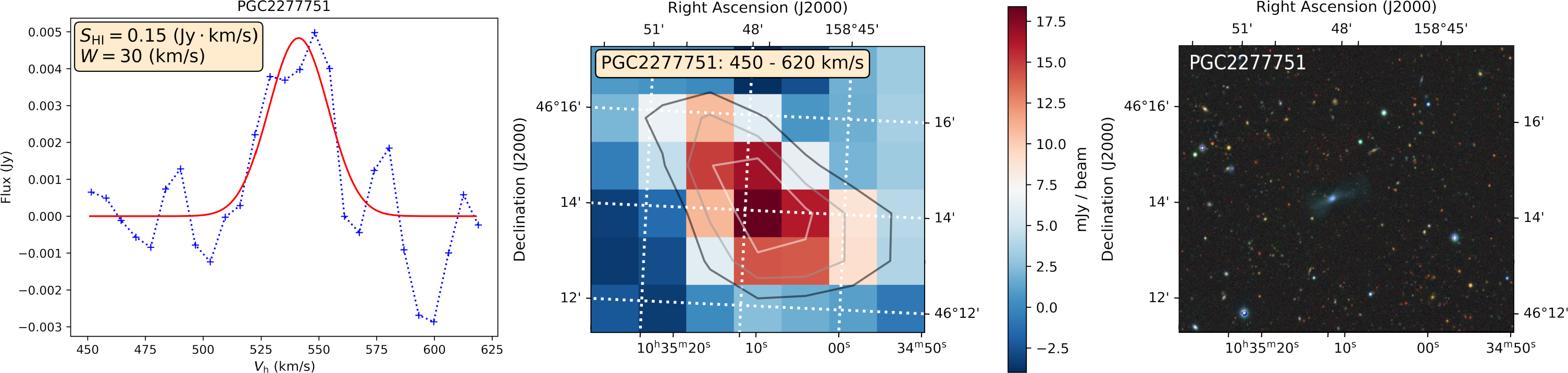}
\end{figure*}
\begin{figure*}[h!]
\centering
    \includegraphics[width=\textwidth]{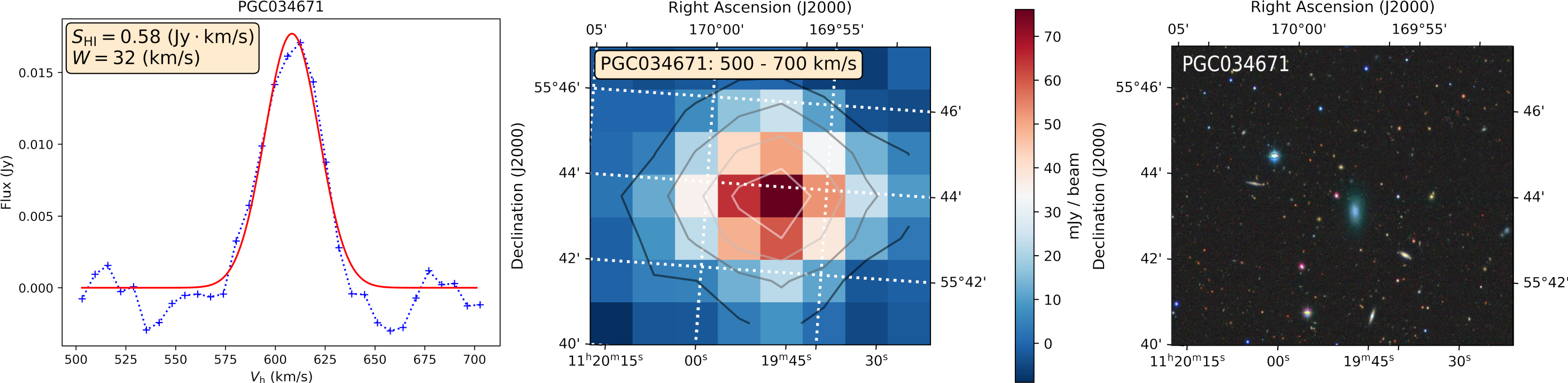}
\end{figure*}
\begin{figure*}[h!]
\centering
    \includegraphics[width=\textwidth]{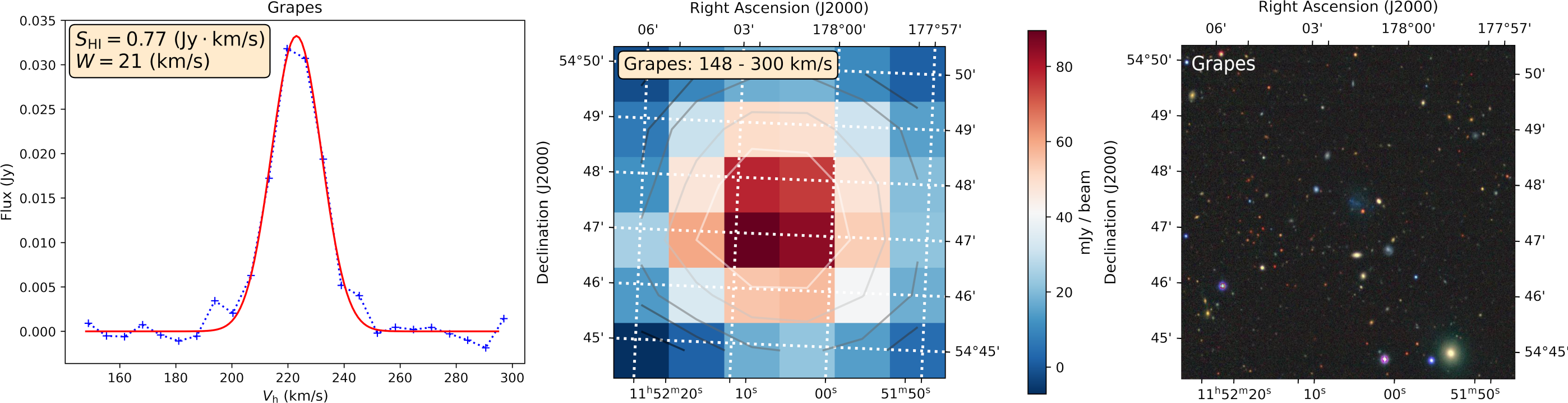}
\end{figure*}
\begin{figure*}[h!]
\centering
    \includegraphics[width=\textwidth]{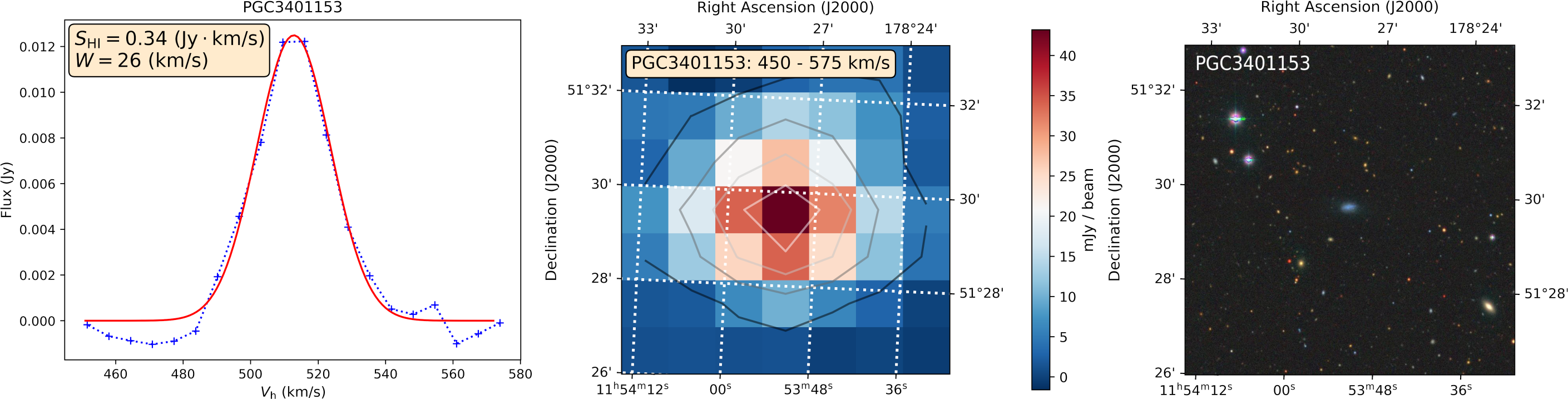}
\end{figure*}
\begin{figure*}[h!]
\centering
    \includegraphics[width=\textwidth]{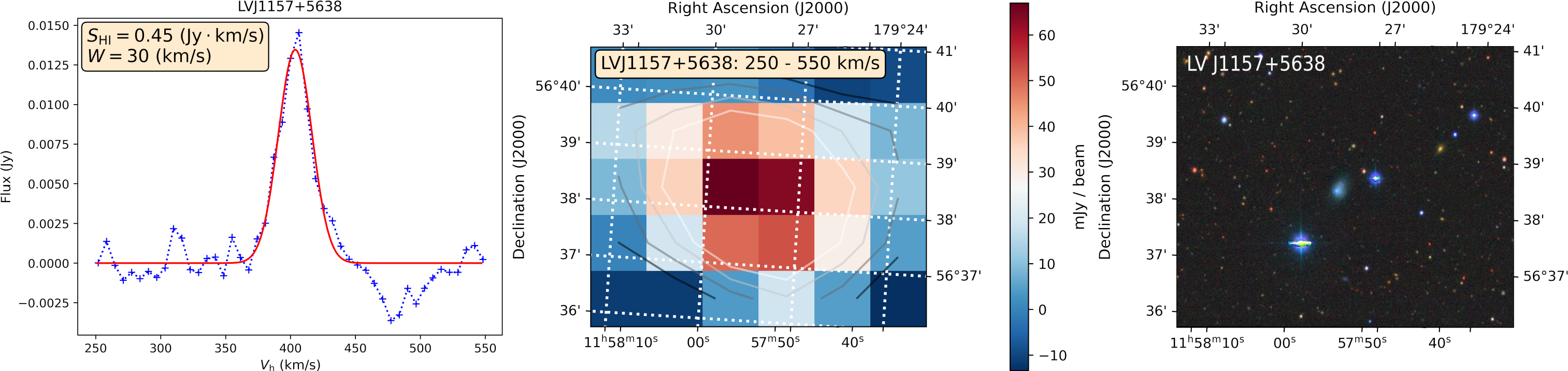}
\end{figure*}
\begin{figure*}[h!]
\centering
    \includegraphics[width=\textwidth]{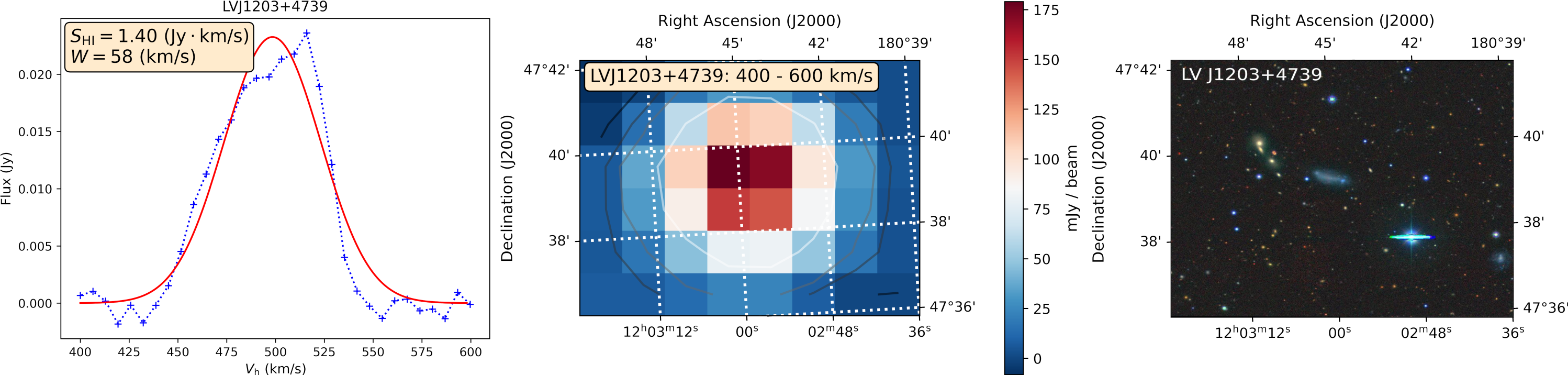}
\end{figure*}
\begin{figure*}[h!]
\centering
    \includegraphics[width=\textwidth]{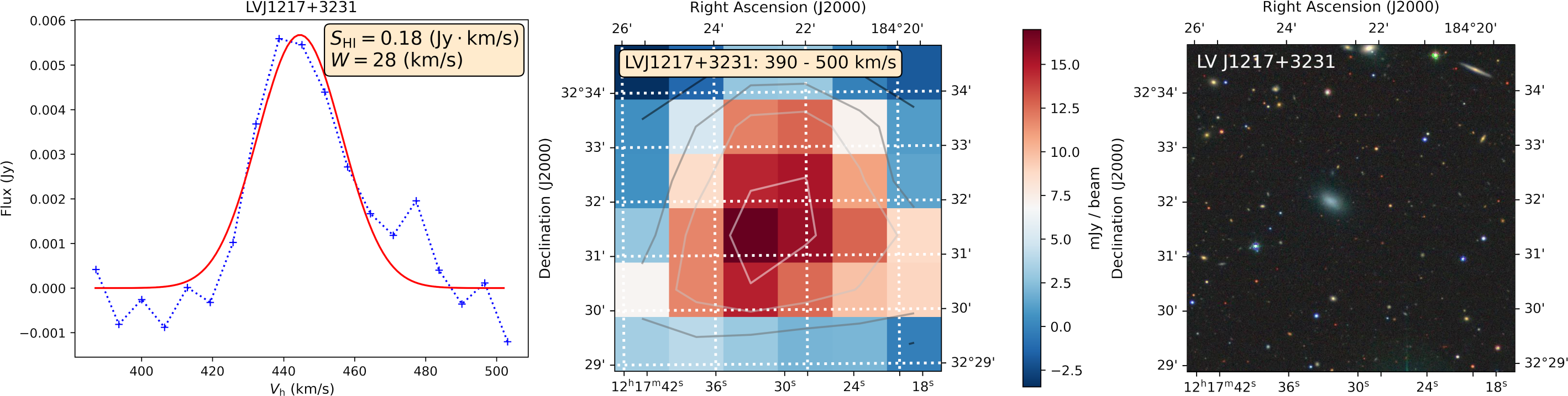}
\end{figure*}
\begin{figure*}[h!]
\centering
    \includegraphics[width=\textwidth]{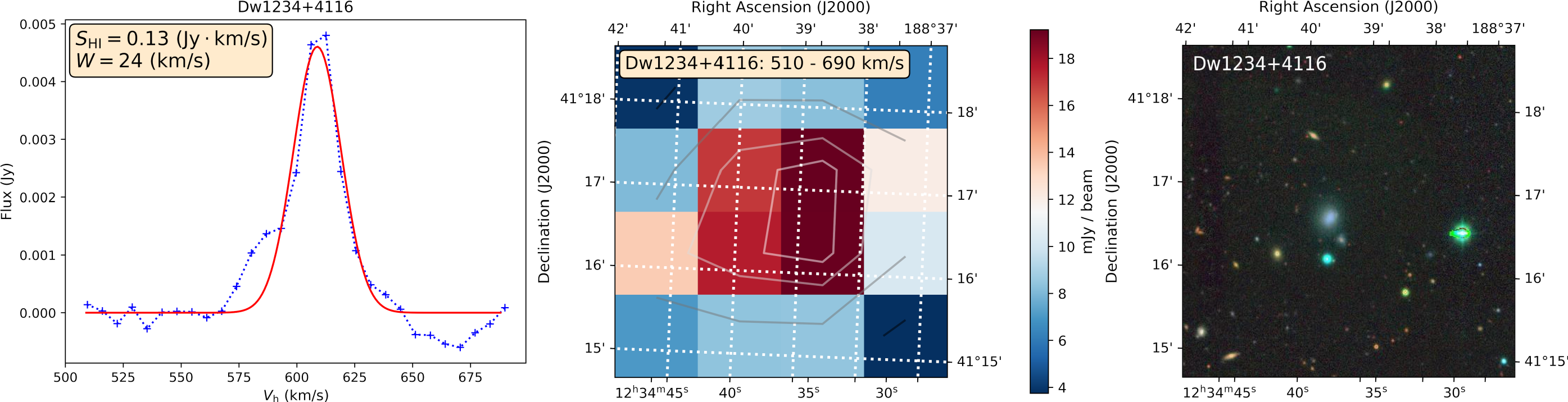}
\end{figure*}
\begin{figure*}[h!]
\centering
    \includegraphics[width=\textwidth]{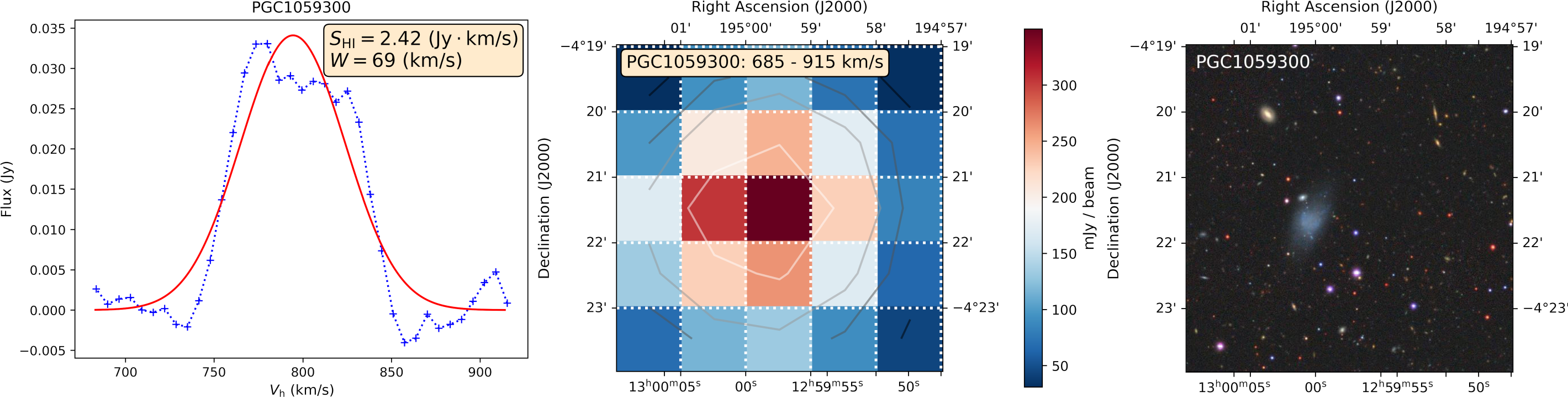}
\end{figure*}
\begin{figure*}[h!]
\centering
    \includegraphics[width=\textwidth]{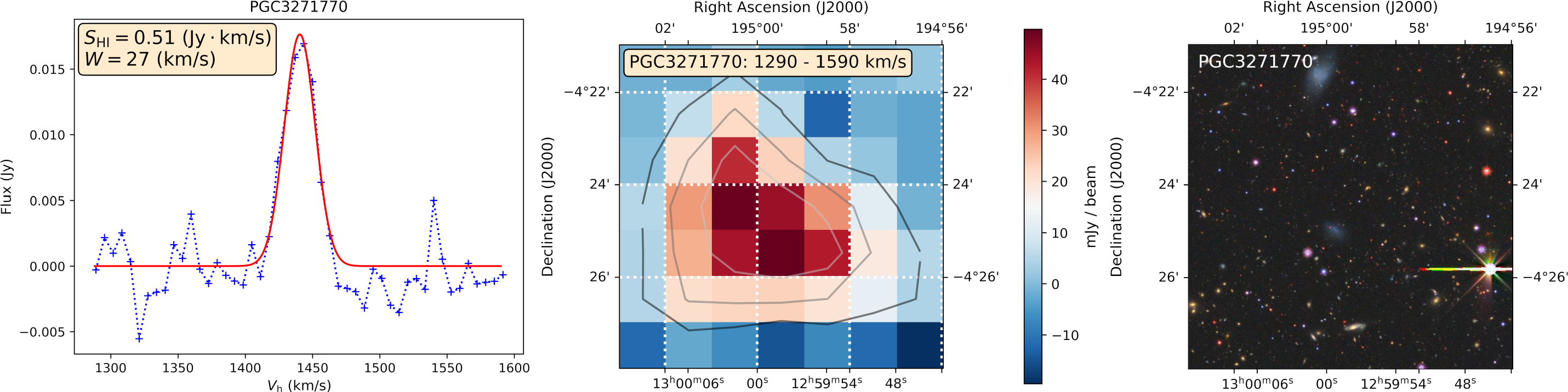}
\end{figure*}
\begin{figure*}[h!]
\centering
     \includegraphics[width=\textwidth]{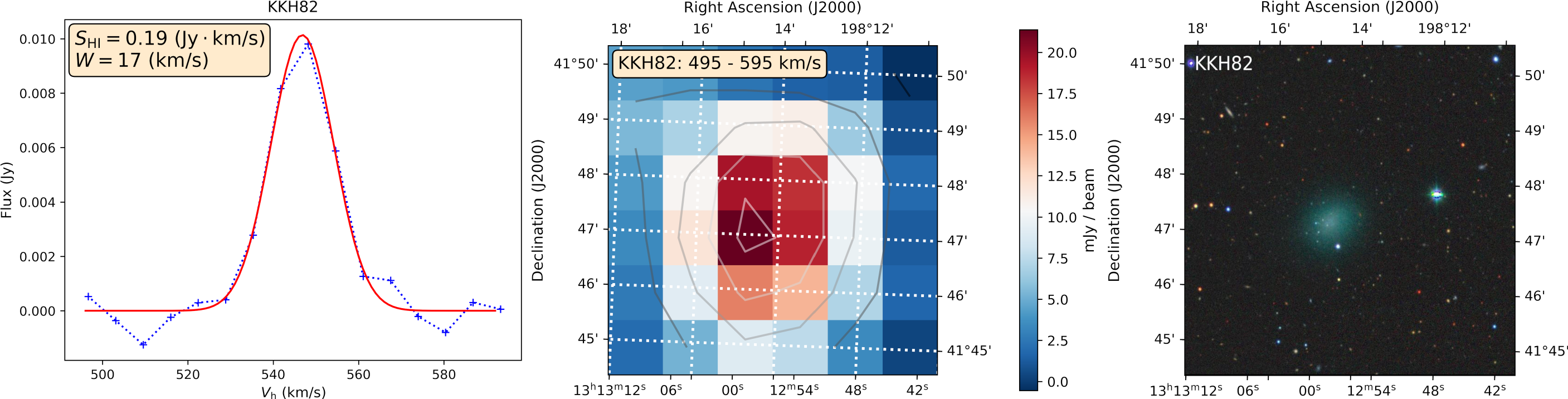}
\end{figure*}
 
\end{appendix}

\end{document}